\documentclass[a4paper]{article}
\usepackage{aqis}
\usepackage{amsfonts}
\usepackage{amsmath}
\usepackage{graphicx}
\usepackage{subfigure}

\begin{document}

\title{
   Resources for Measurement-Based Quantum Carry-Lookahead Adder
}

\author{
  Agung Trisetyarso 
  \affiliation{1}
   \email{trisetyarso07@a8.keio.jp}  
  \and
  Rodney Van Meter 
  \affiliation{2}
  \email{rdv@sfc.wide.ad.jp}
  \and
	Kohei M. Itoh 
  \affiliation{1}
  \email{kitoh@appi.keio.ac.jp}  
}

\address{1}{
  Department of Applied Physics and Physico-Informatics, Keio University, \\
 Yagami Campus, 3-14-1 Hiyoshi, Kohoku-ku, Yokohama-shi, Kanagawa-ken 223-8522, Japan\\
 Phone:+81-45-566-1594 
}
\address{2}{
  Faculty of Environment and Information Studies, Keio University, \\ 
  Shonan Fujisawa Campus,	5322 Endo, Fujisawa-shi, Kanagawa 252-8520, Japan\\
  Phone:+81-466-49-1100
}
\abstract{
We present the design of a quantum carry-lookahead adder using measurement-based quantum computation. QCLA utilizes MBQC`s ability to transfer quantum states in unit time to accelerate addition. The quantum carry-lookahead adder (QCLA) is faster than a quantum ripple-carry adder; QCLA has logarithmic depth while ripple adders have linear depth. QCLA is an order of magnitude faster than a ripple-carry adder when adding registers longer than 100 qubits but requires a cluster state that is an order of magnitude larger. Hand optimization results in a $\approx$ 26$\%$ reduction in spatial resources for the circuit.
}
\maketitle

Measurement-based quantum computation (MBQC) is a new paradigm for implementing quantum algorithms using a quantum cluster state \cite{Raussendorf03}. All gates in the Clifford group, including CNOT, can be performed in one time step via a large number of concurrent measurements. Remarkably, because both wires and SWAP gates are in the Clifford group, MBQC supports long-distance gates in a single time step.  The Toffoli Phase gate can be executed in two time steps, where the measurement basis for the second step is selected depending on previous measurement outcomes. This adaptive process determines the overall performance of many algorithms. 

The performance and requirements for the MBQC QCLA are evaluated. A direct mapping of the circuit model MBQC is given, followed by the optimization for the circuit. The optimization is done by removing the unnecessary wires between the quantum gates and compressing the spacing of active gates.

In order to obtain the performance and resource requirements of the circuit, we have calculated the execution time, circuit area, the number of qubits in the cluster state, and the number of clustering operations required to make that cluster state. The resources are specified in terms of $n$, the length of each of the registers to be added, in qubits. The area is the height of the cluster state multiplied by the width, measured in cluster state lattice sites. The depth of the circuit is the number of time steps required to execute the circuit, counted in qubit measurements.

Our goal in this paper is to minimize execution time while the other three are measures of the cost. The circuit needs to be separated into two type of resources: $\it{first}$, $\textbf{\textit{computational resources}}$ include NOT, CNOT, and Toffoli Phase Gates. $\it{Second}$, $\textbf{\textit{communication resources}}$ are the required for SWAP gate and wires. An optimal circuit is one which contains no communication resources, only computational gates.

\begin{figure}[!ht]
\includegraphics[height=45mm]{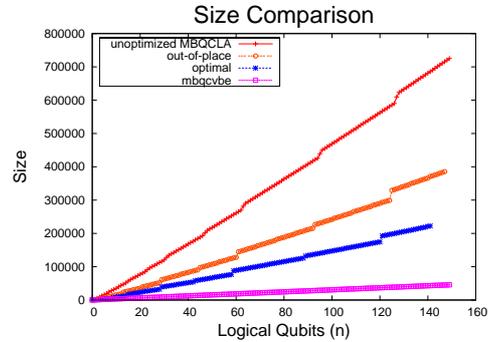}
\caption{\label{comparison1}Size comparison between MBQC ripple-carry and MBQC QCLA. ``+", ``$\diamond$", ``$\star$" and "$\square$" marks are for unoptimized, optimized, and optimal QCLA and ripple-carry circuits.}
\end{figure}
\noindent
In our circuit, the depth is reduced to $\lfloor$log$n$$\rfloor$+$\lfloor$log($n$-1)$\rfloor$+$\lfloor$log$\frac{n}{3}$$\rfloor$+$\lfloor$log$\frac{n-1}{3}$$\rfloor$+14 for QCLA compared to $\approx$O(\textit{n}) for the ripple-carry. However, this circuit costs more in physical resources $\approx$2896$\textit{n}$+64$\textit{n}$$\lfloor$log$_{2}$(\textit{n})$\rfloor$ compared to $\approx$304\textit{n} for the ripple-carry\cite{Raussendorf03}\cite{Draper06}. The optimization of spatial resources in the circuit is $\approx$336$n$+210$w(n)$+14$\lfloor$log$_{2}$$n$$\rfloor$ yielding $\approx$26\% reduction in spatial resources for MBQC QCLA circuit. This optimization results in a near-optimal circuit. 

\paragraph{Acknowledgments}

Supported in part by Grant-in-Aid for Scientific Research by MEXT, Specially Promoted Research No. 18001002 and in part by Special Coordination Funds for Promoting Science and Technology.

\end{document}